\documentclass [12pt,a4paper]{article}
\usepackage{amssymb}
\usepackage{amsthm}
\usepackage{amsmath}
\usepackage{amsfonts}
\usepackage{hyperref}

\newtheoremstyle{mystyle}{}{}{\itshape}{\parindent}{\bfseries}{.}{5mm}{\thmnote{#3}}
\theoremstyle{mystyle}
\newtheorem{theorem}{Theorem}

\newtheorem{corollary}{Corollary}

\newtheorem{definition}{Definition}

\newtheorem{lemma}{Lemma}

\newtheorem{remark}{Remark}

\begin{document}

\title{Construction and Count of Boolean Functions
of an Odd Number of Variables with Maximum Algebraic Immunity
\thanks {This work was supported by the National Natural Science
Foundation of China $($Grant 60373092$)$.}}

\author{Na Li, Wen-Feng Qi\\
\normalsize{Department of Applied Mathematics, Zhengzhou} \\
\normalsize{Information Engineering University} \\
\normalsize{P.O.Box 1001-745, Zhengzhou, 450002,}\\
\normalsize{People's Republic of China}\\
\normalsize{E-mail: mylina{\_}1980@yahoo.com.cn,
wenfeng.qi@263.net}}

\date{}

\maketitle

\begin{abstract}
Algebraic immunity has been proposed as an important property of
Boolean functions. To resist algebraic attack, a Boolean function
should possess high algebraic immunity. It is well known now that
the algebraic immunity of an $n$-variable Boolean function is
upper bounded by $\left\lceil {\frac{n}{2}} \right\rceil $. In
this paper, for an odd integer $n$, we present a construction
method which can efficiently generate a Boolean function of $n$
variables with maximum algebraic immunity, and we also show that
any such function can be generated by this method. Moreover, the
number of such Boolean functions is greater than $2^{2^{n-1}}$.
\end{abstract}


\section{Introduction}

Recently, Algebraic attack has gained a lot of attention in
cryptanalysing stream and block cipher systems
\cite{n01}-\cite{f02}. The study on algebraic attack adds an
important property of Boolean functions to be used in
cryptosystems, which is known as algebraic immunity. Possessing
high algebraic immunity is a necessary requirement for a Boolean
function when used in a cryptosystem. Now, it is known that the
algebraic immunity of an $n$-variable Boolean function is upper
bounded by $\left\lceil {\frac{n}{2}} \right\rceil $ \cite{n02}.

Boolean functions with maximum algebraic immunity are an important
class of Boolean functions, and there is an increasing interest in
construction of such Boolean functions. In \cite{d02}, D. K. Dalai
\emph{et al.} first presented a construction method which can
generate some Boolean functions with maximum algebraic immunity.
This construction provides only one high dimension Boolean
function from a low dimension Boolean function, so it can provide
only a few of such Boolean functions. Then, a construction
\cite{d01} keeping in mind the basic theory of annihilator
immunity was presented. In \cite{l01}, the authors gave three
construction methods which each can get a class of Boolean
functions with maximum algebraic immunity from one such given
function. Several classes of symmetric Boolean functions of an
even number of variables with maximum algebraic immunity were
presented in \cite{a02}. However, the number of symmetric Boolean
functions given by them is small. Moreover, it was showed that
there exists only one symmetric Boolean function (besides its
complement) of an odd number of variables with maximum algebraic
immunity \cite{nl01}. So far, there is no literature which pointed
out that how many on earth such Boolean functions are and how one
can construct an arbitrary such function.

In this paper, for an odd integer $n$, we convert the problem of
finding an $n$-variable Boolean function with maximum algebraic
immunity to the problem of finding a $k\times k$ invertible
submatrix of a $2^{n-1}\times 2^{n-1}$ invertible matrix. Thereby
we present a construction method which can efficiently generate an
$n$-variable Boolean function with maximum algebraic immunity, and
we also show that any such function can be constructed by this
method. Finally, we show that the number of such Boolean functions
is equal to the number of $k\times k$ invertible submatrixes of a
$2^{n-1}\times 2^{n-1}$ invertible matrix, and thus the number of
Boolean functions of an odd number of variables with maximum
algebraic immunity is greater than $2^{2^{n-1}}$.

\section{Preliminaries}
Let $\mathbb{F}_{2}^{n}$ be the set of all $n$-tuples of elements
in the finite field $\mathbb{F}_{2}$. To avoid confusion with the
usual sum, we denote the sum over $\mathbb{F}_{2}$ by $ \oplus $.

A Boolean function of $n$ variables is a mapping from
$\mathbb{F}_{2}^{n}$ to $\mathbb{F}_{2}$. Any Boolean function $f$
of $n$ variables can be uniquely represented as

\begin{equation*}
f(x_1,\ldots,x_n)=a_0 \oplus \sum_{1\leq i\leq n}{a_ix_i}\oplus
\sum_{1\leq i < j \leq n}{a_{i,j}x_ix_j}\oplus \ldots \oplus
a_{1,\ldots,n}x_1x_2\ldots x_n,
\end{equation*}

\noindent where the coefficients
$a_0,a_i,a_{i,j},\ldots,a_{1,\ldots,n} \in \mathbb{F}_{2}$. And
such form of $f$ is called the algebraic normal form (ANF) of $f$.
The algebraic degree, deg$(f)$, is the number of variables in the
highest order term with nonzero coefficient. The Boolean function
$f$ can also be identified by its truth table which is the vector
of length 2$^{n}$ consisting of the function values. The set of
$X\in \mathbb{F}_{2}^{n}$ for which $f(X)=1$ (resp. $f(X)=0$) is
called the onset (resp. offset), denoted by $1_f$ (resp. $0_f$).
The cardinality of $1_f$ is called the Hamming wight of $f$,
denoted by $wt(f)$. We say that an $n$-variable Boolean function
$f$ is balanced if $wt(f)=2^{n-1}$. Let $S=(s_{1},s_{2},\ldots
,s_{n}) \in \mathbb{F}_{2}^{n}$, the Hamming weight of $S$,
denoted by $wt(S)$, is the number of 1's in $\{s_{1},s_{2},\ldots
,s_{n}\}$.

\begin{definition}[Definition 1 {\cite{w01}}]
\label{definition001} For a given $n$-variable Boolean function
$f$, a nonzero $n$-variable Boolean function $g$ is called an
annihilator of $f$ if $f\cdot g=0$, and the algebraic immunity
(AI) of $f$, denoted by AI$(f)$, is the minimum value of $d$ such
that $f$ or $f \oplus 1$\ admits an annihilating function of
degree $d$.

\end{definition}

An important step in the algebraic attack is to find out low
degree annihilators of a Boolean function or its complement. Thus
in order to resist algebraic attacks, neither the Boolean function
nor its complement used in a cryptosystem should have an
annihilator of low degree. That is, the Boolean function should
have high algebraic immunity. In the next section, we will present
a construction method to generate Boolean functions of an odd
number of variables which achieve the maximum algebraic immunity.

\section{Construction and Count}

Let $f$ be a Boolean function of $n$ variables, and
$$
1_f=\{X_1,\ldots,X_{wt(f)}\},0_f=\{X_{wt(f)+1},\ldots,X_{2^{n}}\}.
$$

\noindent It is clear that an $n$-variable Boolean function $g$ is
an annihilator of $f$ if and only if $1_f\subseteq 0_g$. For
$X=(x_1,\ldots,x_n)\in \mathbb{F}_{2}^{n}$, we let

\begin{equation*}
v(X)=(1,x_1,\ldots,x_n,x_1x_2,\ldots,x_{n-1}x_n,\ldots\ldots,x_1\cdot\cdot\cdot
x_{\left\lceil {\frac{n}{2}} \right\rceil
-1},\ldots,x_{\left\lfloor {\frac{n}{2}} \right\rfloor
+2}\cdot\cdot\cdot x_n),
\end{equation*}
which belongs to $\mathbb{F}_{2}^{\sum_{i=0}^{\left\lceil
{\frac{n}{2}} \right\rceil -1}\binom{n}{i}}$. Let $V(1_f)$ be the
$wt(f)\times{\sum_{i=0}^{\left\lceil {\frac{n}{2}} \right\rceil
-1}\binom{n}{i}}$ matrix with  row vectors
$v(X_1),\ldots,v(X_{wt(f)})$ and $V(0_f)$ the
$(2^{n}-wt(f))\times{\sum_{i=0}^{\left\lceil {\frac{n}{2}}
\right\rceil -1}\binom{n}{i}}$ matrix with row vectors
$v(X_{wt(f)+1}),\ldots,v(X_{2^{n}})$.

\begin{lemma}[Lemma 1]
Let $f$ be a Boolean function of $n$ variables. Then
AI$(f)=\left\lceil {\frac{n}{2}} \right\rceil$ if and only if the
ranks of $V(1_f)$ and $V(0_f)$ are both $\sum_{i=0}^{\left\lceil
{\frac{n}{2}} \right\rceil -1}\binom{n}{i}$.
\end{lemma}

\begin{proof}
If there exists a linear relationship among the columns of
$V(1_f)$ (resp. $V(0_f)$), then an annihilator of $f$ (resp.
$f\oplus1$) with degree less than $\left\lceil {\frac{n}{2}}
\right\rceil$ can be found. On the other hand, if there is an
annihilator of $f$ (resp. $f\oplus1$) with degree less than
$\left\lceil {\frac{n}{2}} \right\rceil$, then there must exist a
linear relationship among the columns of $V(1_f)$ (resp.
$V(0_f)$). Therefore, AI$(f)=\left\lceil {\frac{n}{2}}
\right\rceil$ if and only if the ranks of  $V(1_f)$ and $V(0_f)$
are both $\sum_{i=0}^{\left\lceil {\frac{n}{2}} \right\rceil
-1}\binom{n}{i}$.

\end{proof}

Note that for odd integer $n$, $\sum_{i=0}^{\left\lceil
{\frac{n}{2}} \right\rceil -1}\binom{n}{i}=2^{n-1}$, and any
$n$-variable Boolean function with maximum algebraic immunity must
be balanced \cite{d03}. Furthermore, such functions have the
following property.

\begin{lemma}[Lemma 2]\cite{canteaut01}
\label{lemma001} Let odd integer $n=2t+1$, and $f$ be an
$n$-variable balanced Boolean function. If $f$ does not have any
annihilator with degree less than $t+1$, then $f\oplus 1$ has no
annihilator with degree less than $t+1$. Consequently,
AI$(f)=t+1$.
\end{lemma}

\begin{corollary}[Corollary 1]
Let odd integer $n=2t+1$ and  $f$ be an $n$-variable Boolean
function. Then, AI$(f) =t+1$ if and only if $f$ is balanced and
$V(1_f)$ is invertible.

\end{corollary}

\begin{lemma}[Lemma 3]\cite{d01}\cite{a02}
\label{lemma002} Let odd integer $n=2t+1$ and $f$ be an
$n$-variable Boolean function which satisfies

$$
f(X)=\left\{ \begin{array}{*{20}c}
 {a     \quad \mbox{ if }~wt(X) \le t}  \\
 {a \oplus 1\quad \mbox{ if }~wt(X) > t}  \\
\end{array}  \right.,
$$
\noindent where $a \in \mathbb{F}_{2}$, then AI$(f)=t+1$.
\end{lemma}

\begin{remark}[Remark 1]
If $a=1$, we denote the function described in Lemma 3 by $G_n$.
\end{remark}

Let odd integer $n=2t+1$, $F_n$ be a Boolean function of $n$
variables with maximum algebraic immunity (for example,
$F_n=G_n$), and we may let
$$
1_{F_n}=\{Y_1,\ldots,Y_{2^{n-1}}\},0_{F_n}=\{Z_1,\ldots,Z_{2^{n-1}}\}.
$$ Then $V(1_{F_n})$ and $V(0_{F_n})$ are both $2^{n-1}\times 2^{n-1}$
square matrixes, and their row vectors are
$v(Y_1),\ldots,v(Y_{2^{n-1}})$ and $v(Z_1),\ldots,v(Z_{2^{n-1}})$
respectively. By Lemma 1, $V(1_{F_n})$ and $V(0_{F_n})$ are both
invertible matrixes. It is clear that a Boolean function $f$ is
balanced if and only if there exist some integer $0\leq k \leq
2^{n-1}$, integers $1\leq i_1 < \ldots < i_k \leq 2^{n-1}$ and
integers $1\leq j_1 < \ldots < j_k \leq 2^{n-1}$, such that
$$
1_f = \{Z_{i_1},\ldots,Z_{i_k}\} \cup 1_{F_n} \backslash
\{Y_{j_1},\ldots,Y_{j_k}\}
$$ and
$$
0_f = \{Y_{j_1},\ldots,Y_{j_k}\}\cup 0_{F_n} \backslash
\{Z_{i_1},\ldots,Z_{i_k}\} .
$$
So, for some integer $1\leq k \leq 2^{n-1}$, if we can find some
integers $1\leq i_1 < \ldots < i_k \leq 2^{n-1}$ and integers
$1\leq j_1 < \ldots < j_k \leq 2^{n-1}$, such that the
$2^{n-1}\times 2^{n-1}$ matrix with the set of row vectors $
\{v(Z_{i_1}),\ldots,v(Z_{i_k})\} \cup V(1_{F_n}) \backslash
\{v(Y_{j_1}),\ldots,v(Y_{j_k})\}$ is invertible, then by Corollary
1, we can construct a balanced $n$-variable Boolean function
$f_{(i_1,\ldots,i_k;j_1,\ldots,j_k)}(X)$ with maximum algebraic
immunity as follows
\begin{equation}
\label{equation001}
 f_{(i_1,\ldots,i_k;j_1,\ldots,j_k)}(X)=\left\{
\begin{array}{*{20}c}
 {F_n(X)\oplus 1 }\\
 {F_n(X) }
\end{array}
\begin{array}{*{20}c}
{\mbox{if } X\in \{Z_{i_1},\ldots,Z_{i_k},Y_{j_1},\ldots,Y_{j_k}}\}\\
{\mbox{ else }}
\end{array}
\right..
\end{equation}

\noindent This is the core idea of our construction. The following
is a basic conclusion of vector space.

\begin{lemma}[Lemma 4]
Let $U$ be an $m$-dimension vector space with $m\geq2$,
$\{\alpha_1,\ldots,\\\alpha_m\}$ and $\{\beta_1,\ldots,\beta_m\}$
two bases of $U$. Then, for integer $1\leq k \leq m-1$ and
integers $1\leq i_1 < \ldots < i_k \leq m$, there always exist
some integers $1\leq j_1 < \ldots < j_{m-k} \leq m$, such that

$$
\{\alpha_{i_1},\ldots,\alpha_{i_k},\beta_{j_1},\ldots,\beta_{j_{m-k}}\}
$$
is also a base of $U$.
\end{lemma}

\begin{corollary}[Corollary 2]
Let odd integer $n=2t+1$, $F_n$ be a Boolean function of $n$
variables with maximum algebraic immunity and $
1_{F_n}=\{Y_1,\ldots,Y_{2^{n-1}}\},0_{F_n}=\{Z_1,\ldots,Z_{2^{n-1}}\}.
$ Then, for any integer $1\leq k \leq 2^{n-1}-1$ and integers
$1\leq i_1 < \ldots < i_k \leq 2^{n-1}$, there always exist some
integers $1\leq j_1 < \ldots < j_{k} \leq 2^{n-1}$, such that
AI$(f_{(i_1,\ldots,i_k;j_1,\ldots,j_k)})=t+1$, where
$f_{(i_1,\ldots,i_k;j_1,\ldots,j_k)}$ is defined by $(1)$.

\end{corollary}

\begin{proof}
Since $V(1_{F_n})$ and $V(0_{F_n})$ are both invertible, then
$\{v(Y_1),\ldots,v(Y_{2^{n-1}})\}$ and
$\{v(Z_1),\ldots,v(Z_{2^{n-1}})\}$ are two bases of
$2^{n-1}$-dimension vector space $\mathbb{F}_{2}^{2^{n-1}}$. By
Lemma 4, for any integer $1\leq k \leq 2^{n-1}-1$ and integers
$1\leq i_1 < \ldots < i_k \leq 2^{n-1}$, there always exist some
integers $1\leq j_1 < \ldots < j_{k} \leq 2^{n-1}$, such that $
\{v(Z_{i_1}),\ldots,v(Z_{i_k})\} \cup V(1_{F_n}) \backslash
\{v(Y_{j_1}),\ldots,v(Y_{j_k})\}$ is a base of
$\mathbb{F}_{2}^{2^{n-1}}$. That is, the matrix with the set of
row vectors $ \{v(Z_{i_1}),\ldots,v(Z_{i_k})\} \cup V(1_{F_n})
\backslash \{v(Y_{j_1}),\ldots,v(Y_{j_k})\}$ is invertible.
Therefore AI$(f_{(i_1,\ldots,i_k;j_1,\ldots,j_k)})=t+1$.
\end{proof}

Next, we show how to find those $1\leq j_1 < \ldots < j_{k} \leq
2^{n-1}$ for given $1\leq k \leq 2^{n-1}-1$ and $1\leq i_1 <
\ldots < i_k \leq 2^{n-1}$.

\bigskip
\textbf{A useful matrix $W(F_n)$}. Let odd integer $n=2t+1$, $F_n$
be a Boolean function of $n$ variables with maximum algebraic
immunity and $
1_{F_n}=\{Y_1,\ldots,\\Y_{2^{n-1}}\},0_{F_n}=\{Z_1,\ldots,Z_{2^{n-1}}\}.
$ Set
$$
W(F_n)=V(0_{F_n})V(1_{F_n})^{-1}.
$$ Then $W(F_n)$ is a
$2^{n-1} \times 2^{n-1} $ invertible matrix. Denote the $2^{n-1}$
row vectors of $W(F_n)$ by $w(F_n)_1,\ldots,w(F_n)_{2^{n-1}}$.
From the definition of $W(F_n)$, we have
$V(0_{F_n})=W(F_n)V(1_{F_n})$, that is,
$$
\left(%
\begin{array}{c}
  v(Z_1) \\
  v(Z_2)\\
  \ldots \\
  v(Z_{2^{n-1}})\\
\end{array}%
\right)=\left(%
\begin{array}{c}
  w(F_n)_1 \\
  w(F_n)_2\\
  \ldots \\
  w(F_n)_{2^{n-1}}\\
\end{array}%
\right)\left(%
\begin{array}{c}
  v(Y_1) \\
  v(Y_2)\\
  \ldots \\
  v(Y_{2^{n-1}})\\
\end{array}%
\right).
$$

\bigskip
The following theorem is one of our main result.

Let $W(F_n)_{(i_1,\ldots,i_k)}$ denote the $k\times 2^{n-1}$
matrix with row vectors $w(F_n)_{i_1},\ldots,\\w(F_n)_{i_k}$ and
$W(F_n)_{(i_1,\ldots,i_k;j_1,\ldots,j_k)}$ denote the $k\times k$
matrix with column vectors equal to the $j_1th,\ldots,j_kth$
columns of $W(F_n)_{(i_1,\ldots,i_k)}$.

\begin{theorem}[Theorem 1]
Let odd integer $n=2t+1$, $F_n$ be a Boolean function of $n$
variables with maximum algebraic immunity and $
1_{F_n}=\{Y_1,\ldots,Y_{2^{n-1}}\},0_{F_n}=\{Z_1,\ldots,Z_{2^{n-1}}\}.
$ Then, the set
\begin{equation*}
\begin{split}
\{&f_{(i_1,\ldots,i_k;j_1,\ldots,j_k)}|k=0,\ldots,2^{n-1},1\leq
i_1 <\ldots < i_k \leq 2^{n-1},\\&1\leq j_1 <\ldots < j_k \leq
2^{n-1}, W(F_n)_{(i_1,\ldots,i_k;j_1,\ldots,j_k)}~
 \text{is invertible}\}
 \end{split}
\end{equation*}
consists of all $n$-variable Boolean functions with maximum
algebraic immunity, where $f_{(i_1,\ldots,i_k;j_1,\ldots,j_k)}$ is
defined by (1) and $W(F_n)_{(i_1,\ldots,i_k;j_1,\ldots,j_k)}$ is
defined as above.
\end{theorem}

\begin{proof}

Since an $n$-variable Boolean function $f$ with maximum algebraic
immunity must be balanced, then $f$ must be of the form
$f_{(i_1,\ldots,i_k;j_1,\ldots,j_k)}$. Denote the remaining
elements of $1_{F_{n}}$ (resp. $0_{F_{n}}$) excluding
$Y_{j_1},\ldots,Y_{j_k}$ (resp. $Z_{i_1},\ldots,Z_{i_k}$) by
$Y_{k+1}^{'},\ldots,Y_{2^{n-1}}^{'}$ (resp.
$Z_{k+1}^{'},\ldots,Z_{2^{n-1}}^{'}$). Then
$V(1_{f_{(i_1,\ldots,i_k;j_1,\ldots,j_k)}})$ is a $2^{n-1}\times
2^{n-1}$ matrix with row vectors
$$
v(Z_{i_1}),\ldots,v(Z_{i_k}),v(Y_{k+1}^{'}),\ldots,v(Y_{2^{n-1}}^{'}).
$$

By Corollary 1, AI$(f_{(i_1,\ldots,i_k;j_1,\ldots,j_k)})=t+1$ if
and only if $V(1_{f_{(i_1,\ldots,i_k;j_1,\ldots,j_k)}})$ is
invertible. Therefore, it is sufficient to prove that
$V(1_{f_{(i_1,\ldots,i_k;j_1,\ldots,j_k)}})$ is invertible if and
only if $W(F_n)_{(i_1,\ldots,i_k;j_1,\ldots,j_k)}$ is invertible.

Let $M$ denote the $k\times (2^{n-1}-k)$ matrix with column
vectors equal to the remaining columns of
$W(F_n)_{(i_1,\ldots,i_k)}$ which is defined as above, such that

$$
\left(
\begin{array}{c}
 v(Z_{i_1})\\
  \ldots \\
  v(Z_{i_k})\\
 \end{array}
\right)=W(F_n)_{(i_1,\ldots,i_k;j_1,\ldots,j_k)}\left(%
\begin{array}{c}
  v(Y_{j_1})\\
 \ldots \\
  v(Y_{j_k})\\
\end{array}%
\right)\oplus M\left(%
\begin{array}{c}
  v(Y_{k+1}^{'}) \\
  \ldots\\
  v(Y_{2^{n-1}}^{'})\\
\end{array}%
\right).
$$

Then, we have

\begin{equation}
\left(
\begin{array}{c}
 v(Z_{i_1})\\
  \ldots \\
  v(Z_{i_k})\\
  v(Y_{k+1}^{'}) \\
  \ldots\\
  v(Y_{2^{n-1}}^{'})\\
\end{array}
\right)=\left(
\begin{array}{cccc}
  W(F_n)_{(i_1,\ldots,i_k;j_1,\ldots,j_k)} &  & M &  \\
    & 1 &   &   \\
  0 &  & \ldots &  \\
  &  &  & 1 \\
\end{array}
\right)\left(
\begin{array}{c}
 v(Y_{j_1})\\
  \ldots \\
  v(Y_{j_k})\\
  v(Y_{k+1}^{'}) \\
  \ldots\\
  v(Y_{2^{n-1}}^{'})\\
\end{array}
\right).
\end{equation}

\noindent From (2), it is obvious that
$V(1_{f_{(i_1,\ldots,i_k;j_1,\ldots,j_k)}})$ is invertible if and
only if the matrix
\begin{equation}
\label{equation004}
\left(%
\begin{array}{cccc}
  W(F_n)_{(i_1,\ldots,i_k;j_1,\ldots,j_k)} &  & M &  \\
    & 1 &   &   \\
  0 &  & \ldots &  \\
  &  &  & 1 \\
\end{array}%
\right)
\end{equation}
is invertible. Further, the matrix (3) is invertible if and only
if $W(F_n)_{(i_1,\ldots,i_k;j_1,\ldots,j_k)}$ is invertible. Thus
the proof is completed.

\end{proof}

\begin{remark}[Remark 2]
Since $W(F_n)$ is a $2^{n-1}\times 2^{n-1}$ invertible matrix, for
any integer $1\leq k \leq 2^{n-1}-1$ and integers $1\leq i_1
<\ldots < i_k \leq 2^{n-1}$, the rank of the $k\times 2^{n-1}$
matrix $W(F_n)_{(i_1,\ldots,i_k)}$ is $k$, which means there must
exist some integers $1\leq j_1 <\ldots < j_k \leq 2^{n-1}$ (we
note that there may exist many groups of these integers) such that
$W(F_n)_{(i_1,\ldots,i_k;j_1,\ldots,j_k)}$ is invertible. We can
also derive Corollary 2 by this fact.
\end{remark}

In order to efficiently generate Boolean functions of an odd
number of variables with maximum algebraic immunity, we should
choose those $F_n$ such that $W(F_n)$ can be efficiently obtained.
We note that $G_n$ is such a function. Now, we explain how to
obtain the matrix $W(G_n)$. We denote the elements of $1_{G_n}$
and $0_{G_n}$ by some special symbols. Let
$Y_{(b_1,\ldots,b_i)}=(y_1,\ldots,y_n) \in 1_{G_{n}}$, where
$1\leq b_1<\ldots<b_i\leq n$. The symbol $Y_{(b_1,\ldots,b_i)}$
means that $wt(Y_{(b_1,\ldots,b_i)})=i$ and $y_s=1$ only for
$s=b_1,\ldots,b_i$. Let $Y_{(0)}$ denote $(0,\ldots,0)$.
Similarly, let $Z_{(a_1,\ldots,a_l)}=(z_1,\ldots,z_n) \in
0_{G_{n}}$, where $1\leq a_1<\ldots<a_l\leq n$. The symbol
$Z_{(a_1,\ldots,a_l)}$ means that $wt(Z_{(a_1,\ldots,a_l)})=l$ and
$z_s=1$ only for $s=a_1,\ldots,a_l$. It is clear that
$wt(Z_{(a_1,\ldots,a_l)})\geq t+1$ since $Z_{(a_1,\ldots,a_l)}\in
0_{G_{n}}$. Then, the vector $v(Z_{(a_1,\ldots,a_l)})$ can be
expressed as a linear combination of the row vectors of
$V(1_{G_n})$ as follows.

\begin{equation}
\label{equation002}
\begin{split}
v(Z_{(a_1,\ldots,a_l)})&=c_0\sum _{\{b_1,\ldots,b_t\}\subseteq
\{a_1,\ldots,a_l\}}{v(Y_{(b_1,\ldots,b_t)})}\\&\oplus c_1\sum
_{\{b_1,\ldots,b_{t-1}\}\subseteq
\{a_1,\ldots,a_l\}}{v(Y_{(b_1,\ldots,b_{t-1})})}\\&\oplus c_2\sum
_{\{b_1,\ldots,b_{t-2}\}\subseteq
\{a_1,\ldots,a_l\}}{v(Y_{(b_1,\ldots,b_{t-2})})}\oplus \ldots
\\&\oplus c_i\sum _{\{b_1,\ldots,b_{t-i}\}\subseteq
\{a_1,\ldots,a_l\}}{v(Y_{(b_1,\ldots,b_{t-i})})}\oplus \ldots
\\&\oplus c_{t-1}\sum _{\{b_1\}\subseteq
\{a_1,\ldots,a_l\}}{v(Y_{(b_1)})} \oplus c_{t}v(Y_{(0)}),
\end{split}
\end{equation}

\noindent where
$$
c_0=1;
$$
$$
c_i=1\oplus c_0\binom{l}{i}\oplus c_1\binom{l}{i-1}\oplus \ldots
\oplus c_{i-1}\binom{l}{1}.
$$

\noindent From (4), we get the corresponding row vector of
$W(G_n)$. And the other row vectors of $W(G_n)$ can also be
obtained by this method.

Now, we derive our important result.

\bigskip
\textbf{Construction.} Let odd integer $n=2t+1$,
$1_{G_n}=\{Y_1,\ldots,Y_{2^{n-1}}\},0_{G_n}=\{Z_1,\ldots,Z_{2^{n-1}}\}.
$ To find a Boolean function of $n$ variables with maximum
algebraic immunity, what one has to do is the following steps.

Step 1: Select randomly an integer $1\leq k \leq 2^{n-1}-1$ and
$k$ integers $1\leq i_1 < \ldots < i_k \leq 2^{n-1}$;

Step 2: Using Gauss elimination on the column vectors of
$W(G_n)_{(i_1,\ldots,i_k)}$, find a group of integers $1\leq j_1 <
\ldots < j_k \leq 2^{n-1}$, such that the $j_1th,\ldots,j_kth$
column vectors of $W(G_n)_{(i_1,\ldots,i_k)}$ are linear
independent.

We construct the Boolean function
$f_{(i_1,\ldots,i_k;j_1,\ldots,j_k)}$ as follows.
\begin{equation}
\label{equation001}
 f_{(i_1,\ldots,i_k;j_1,\ldots,j_k)}(X)=\left\{
\begin{array}{*{20}c}
 {G_n(X)\oplus 1 }\\
 {G_n(X) }
\end{array}
\begin{array}{*{20}c}
{\mbox{if } X\in \{Z_{i_1},\ldots,Z_{i_k},Y_{j_1},\ldots,Y_{j_k}}\}\\
{\mbox{ else }}
\end{array}
\right..
\end{equation}
Then $f_{(i_1,\ldots,i_k;j_1,\ldots,j_k)}$ achieves the maximum
algebraic immunity $t+1$.
\begin{remark}[Remark 3]
(i) By Theorem 1, it is clear that any Boolean function of an odd
number of variables with maximum algebraic immunity can be
constructed by our method.

(ii)Since AI$(f)$=AI$(f\oplus 1)$, The range of value of $k$ in
Step 1 only needs to be $1\leq k \leq 2^{n-2}$.

(iii)For a small $k$, one can efficiently generate an $n$-variable
Boolean function with maximum algebraic immunity. For example,
when $k=1$, we first select randomly an integer $1\leq i \leq
2^{n-1}$ according to Step 1. then according to Step 2, we can
select any integer $1\leq j \leq 2^{n-1}$ such that the $jth$
element of $w(G_n)_i$ is $1$. Thus we generate a Boolean function
$f_{(i;j)}$.

\end{remark}

Finally, we get a result on the count of Boolean functions of an
odd number of variables with maximum algebraic immunity.

\begin{theorem}[Theorem 2]
Let $n$ be an odd integer, then the number of $n$-variable Boolean
functions with maximum algebraic immunity is equal to the number
of $k\times k$ invertible submatrixes of $W(G_n)$. Further, it is
greater than $2^{2^{n-1}}$.
\end{theorem}

\begin{proof}
It is clear that for different groups of integers
$(i_1,\ldots,i_k;j_1,\ldots,j_k)$, the Boolean functions defined
by (5) are different.

By Theorem 1, the first conclusion is obvious. By Corollary 2 and
Remark 3, it is clear that the number of $n$-variable Boolean
functions with maximum algebraic immunity is greater than

$$
\binom{2^{n-1}}{0}+\binom{2^{n-1}}{1} + \ldots +\binom
{2^{n-1}}{2^{n-1}}=2^{2^{n-1}}.
$$

\end{proof}

\section{Conclusion}

In this paper, we present a construction method which can
efficiently generate a Boolean function of an odd number of
variables which possesses maximum algebraic immunity, and we show
that any such function can be generated by this method. Based on
the construction, we show that the number of this kind of Boolean
functions is greater than $2^{2^{n-1}}$. This value is great
enough to reveal that this kind of Boolean functions are numerous.
There are some other problems worth studying. For example, how to
construct and count Boolean functions of an even number of
variables with maximum algebraic immunity, how to construct and
count Boolean functions with maximum algebraic immunity keeping in
mind of other cryptographic properties such as nonlinearity,
propagation and resiliency.

\end{document}